\documentclass[pra]{revtex4}
\usepackage{graphicx}
\usepackage{indentfirst}
\usepackage{dcolumn}
\usepackage{amsfonts}
\usepackage{amsmath}
\usepackage{bm}
\newcommand{\ket}[1]{\left|#1\right\rangle}
\newcommand{\bra}[1]{\left\langle#1\right|}
\newcommand{\bgeq}{\begin{equation}}
\newcommand{\edeq}{\end{equation}}
\newcommand{\bgeqn}{\begin{eqnarray}}
\newcommand{\edeqn}{\end{eqnarray}}
\newcommand{\tr}[1]{{\rm Tr\{ #1 \}}}
\newtheorem{theorem}{Proposition}

\newtheorem{corollary}[theorem]{Corollary}

\begin{document}

\title{Complementarity of information sent via different bases}

\author{Shengjun Wu$^{1,2}$}
\author{Sixia Yu$^{2,3}$}
\author{Klaus M{\o}lmer$^1$}%
\affiliation{
$^1$Lundbeck Foundation Theoretical Center for Quantum System Research, \\
Department of Physics and Astronomy, Aarhus University,
DK-8000 Aarhus C, Denmark
\\
$^2$Hefei National Laboratory for Physical Sciences at Microscale and Department of Modern Physics, \\
 University of Science and Technology of China, Hefei, Anhui 230026, P. R.
 China \\
$^3$Centre for Quantum Technologies and Physics Department, \\
National University of Singapore, 2 Science Drive 3, Singapore 117542, Singapore
}

\date{\today}

\begin{abstract}
We discuss quantitatively the complementarity of information transmitted by a quantum system prepared
in a basis state in one out of several different mutually unbiased
bases (MUBs).  We obtain upper bounds on the information available to a receiver
who has no knowledge of which MUB was chosen by the sender.
These upper bounds imply a complementarity of information encoded via different
MUBs and ultimately ensure the security in quantum key distribution protocols.
\end{abstract}

\maketitle

\section{Introduction}

In this paper we consider communication between a sender, Alice, who prepares a quantum system in
a given state and a receiver, Bob, who receives the quantum system after its passage through
a perfect quantum channel. The quantum system is a qudit with a $d$-dimensional Hilbert space,
and if Alice and Bob agree to use a fixed basis for the communication, each quantum system,
prepared in one of the basis states, can thus unambiguously encode a message in the form of an
integer value between 1 and $d$. Quantum mechanics offers the possibility to prepare superposition states,
who can themselves be viewed as basis vectors in a different basis, and from the possibility to use different bases
follows a number of properties and possibilities, e.g.,
to use quantum systems for secure quantum key distribution \cite{BB84,ekert91,BBM92,Bruss98,BG98,Bennett92}.

A large amount of work has been devoted to present security proofs for quantum key distribution
\cite{Mayers98,LC98,SP00}
and to characterize other information theoretical properties, such as the channel capacity of a quantum transmission
channel. It is thus an ongoing effort in quantum information science to investigate and quantify the interplay between
quantum physical properties such as complementarity and uncertainty relations, and information theoretic concepts
such as Shannon entropy and mutual information.
In this paper we shall present a number of results concerning the information that one or several recipients can
obtain about messages sent by a source, Alice, making use of $d$-dimensional quantum systems.
We present some rigid results and some related conjectures in the form of upper bounds on the mutual information
between the sender and receiver about messages sent with different bases.

The paper is organized as follows. In section II, we start with an
overview of the problem, and we present the notation and assumptions
that will be used in the rest of the paper. In section III, we
discuss the complementarity of the information sent via different
mutually unbiased bases (MUBs). First, upper bounds are obtained for
the sum of the information about the messages sent via two MUBs with
both equal and different prior probabilities on basis states. Next,
the case of $M$ MUBs is considered, and upper bounds are derived for
both equal and unequal prior probabilities. We conjecture that the
upper bounds for equal prior probabilities also hold for unequal
prior probabilities. In section IV, we give a detailed discussion of
the two-dimensional (qubit) case, and both analytical and numerical
results are presented. In section V, our results on the
complementarity of the information sent via different MUBs are used
to derive quantitative results that ensure the security of quantum
key distribution protocols. A conclusion is given in section VI.

\section{Notation and assumptions}

Suppose that there are $M$ mutually unbiased $d$-dimensional bases, $\{ \ket{i_m} \}$,
with $1\leq i \leq d$ and $ 1 \leq m \leq M $, that Alice can choose from in her preparation of a quantum system,
and that she will send multiple copies of the quantum system prepared in the basis functions for one of these bases.
MUBs \cite{Ivanovic81,wootters89,KR03,Pittenger04} are defined by the property that the squared overlap between a basis function in one basis
and all basis states in the other bases are identical, and hence the detection of a particular basis state does
not give away any information about the state if it was prepared in another basis.

Alice chooses, say, the $m$th MUB $\{ \ket{i_m} | i=1,\cdots,d \}$, and prepares
the qudit in one of the $d$ basis states according to the prior probabilities
$p^{(m)}_i$ ($p^{(m)}_i \geq 0$, $\sum_{i=1}^d p^{(m)}_i =1$). For convenience, we suppose $p^{(m)}_1\geq p^{(m)}_2 \cdots \geq p^{(m)}_d$.
As illustrated in Fig. 1., she sends the qudit, and Bob, who does not know which basis Alice has chosen, performs
a measurement on the quantum system. We will assume the most general form for
Bob's measurement, i.e., it is characterized by positive operator valued measure (POVM)
elements $\{ M_s \}$, with $M_s \geq 0$ and $\sum_{s=1}^D M_s =I$.
\begin{figure}
\begin{center}
\includegraphics[angle=0,height=1.2in]{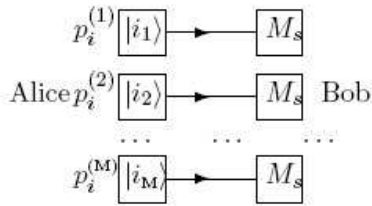}
\end{center}
\caption{Illustration of transmission of a quantum system between Alice and Bob. Each line indicates Alice's use of one particular basis with given basis vector probabilities, and the subsequent
POVM detection by Bob, who has no prior knowledge of the basis chosen by Alice.}
\end{figure}

We now calculate the amount of information available to Bob about the message, i.e.,
the basis vector sent by Alice, without knowing that Alice has chosen a definite, say,
the $m$th MUB. Let $p^{(m)}_{s|i}$ denote the probability that Bob obtains the result $s$, conditional on
the state $\ket{i_m}$ sent by Alice. We have $p^{(m)}_{s|i} = \bra{i_m} M_s \ket{i_m}$.
The joint probability that
Alice sends the state $ \ket{i_m} $ and Bob obtains the result $s$ is given by
\bgeq
p^{(m)}_{is} = p^{(m)}_i p^{(m)}_{s|i}= p^{(m)}_i \bra{i_m} M_s \ket{i_m} .
\edeq
The total probability that Bob gets a result $s$ is thus given by
\bgeq
p^{(m)}(s)=\sum_{i=1}^d p^{(m)}_{is} = \sum_{i=1}^d p^{(m)}_i \bra{i_m} M_s \ket{i_m}.
\label{probBs}
\edeq
Conditioned on the result $s$ obtained
by Bob, the probability that the state sent by Alice is $\{ \ket{i_m} \}$, is therefore given by
\bgeq
p^{(m)}_{i|s}=p^{(m)}_{is}/p^{(m)}(s)= \frac{p^{(m)}_i \bra{i_m} M_s \ket{i_m}}{\sum_{i=1}^d p^{(m)}_i \bra{i_m} M_s \ket{i_m}} .
\label{probigivens}
\edeq
The $m$th information (the information encoded via the $m$th MUB) Bob can obtain is given by the mutual information,
\bgeq
I_m = H\{ p^{(m)}_i;i \} - \sum_{s=1}^D p^{(m)}(s) H\{p^{(m)}_{i|s}; i\}, \label{mthinfor}
\edeq
where $H\{ q_i;i \}$ denotes the entropy $- \sum_{i} q_i \log_2 q_i$ for a probability
distribution $\{ q_i \}$.

Bob can vary his measurement strategy by varying his set of POVM elements. However, he only obtains the maximal information encoded via a particular MUB, by choosing the appropriate projection operators on that basis and in that case he can obtain no information encoded via a different MUB.  We will express this complementarity in terms of upper bounds on the sum of information encoded via different MUBs.

\section{Upper bounds on information encoded via different mutually unbiased bases}

We first consider the case of two mutually unbiased bases, say, $m=1,2$.

\begin{theorem}. \label{the11}
If the prior probabilities on basis states in two MUBs are equal $p_i^{(1)}=p_j^{(2)}=1/d$,
the sum of the information about messages sent in the two bases, available to an observer who does not know
which basis is used, has the upper limit
\bgeq
I_1 + I_2 \leq \log_2 d . \label{2mubseq1}
\edeq
\end{theorem}

{\bf Proof.} Since $p^{(1)}_i =p^{(2)}_i =1/d$, we have $H\{ p^{(1)}_i;i \}=H\{ p^{(2)}_i;i \}= \log_2 d$. From (\ref{probBs}) we also have
\bgeq
p^{(1)}(s)=p^{(2)}(s)= \frac{1}{d} \tr{M_s} \equiv p(s) .
\edeq
Therefore from (\ref{mthinfor}) we have
\bgeq
I_1 +I_2 = 2 \log_2 d - \sum_{s=1}^D p(s) \left( H\{p^{(1)}_{i|s};i\} + H\{p^{(2)}_{i|s};i \} \right) .
\edeq
Now we use a duality between measurements and ensembles \cite{ZB04,griffiths05,wu07} and formally define a state $\rho_s \equiv M_s / \tr{M_s}$.
Rather than Alice sending a state to Bob, we consider the opposite situation where Bob prepares the qudit in one of the states $\rho_s$ according to prior probabilities $p(s) =\frac{1}{d} \tr{M_s} $
and sends the quantum system to Alice. In this reversed picture, Alice may choose one of the MUBs and
perform a projective measurement onto this basis. We then recognize the sum of the terms
inside the bracket in the above equation as the entropic uncertainty for a
state $\rho_s$, which
is known to have a lower bound \cite{maassen88},
\bgeq
H\{p^{(1)}_{i|s};i\} + H\{p^{(2)}_{i|s};i\} \geq \log_2 d .
\label{entropicuncertainty2}
\edeq
Therefore
\bgeq
I_1 +I_2 \leq  2\log_2 d - \sum_{s=1}^D p(s) \log_2 d =\log_2 d
\edeq
which completes the proof of proposition \ref{the11}.

It should be pointed out that our proposition \ref{the11} can also be derived from
inequality (78) in \cite{hall97},  where the information inequality for the case of two source ensembles with
the same ensemble density operator is derived via a more general form of the duality, the so-called {\sl source duality}.
The complementarity in the dual problem is referred to as an {\sl information exclusion relation} in \cite{hall95}.
The duality based argument is not applicable when we discuss the following general case.

When the prior probabilities used by Alice to choose the basis state within the MUBs are different, we can prove
the following weaker upper bounds on the information obtained by Bob about the two bases.

\begin{theorem}.  \label{the12}
For two qudit MUBs, in which the basis states are chosen according to probabilities $p_i^{(1)},\ p_i^{(2)}$, in decreasing order with $i=1, ... ,d$, we can prove that
\bgeqn
\frac{1}{dp_1^{(1)}} I_1  + \frac{1}{dp_1^{(2)}} I_2 \leq \log_2 d  , \label{2mubgeneral1} \\
I_1 + I_2 \leq d p_{max}\log_2 d  . \label{2mubgeneral}
\edeqn
Here $p_{max}= max \{ p_1^{(1)}, p_1^{(2)} \}$ is the maximal value over all prior probabilities in both MUBs, and therefore
(\ref{2mubgeneral1}) is stronger than (\ref{2mubgeneral}).
\end{theorem}

{\bf Proof.}
For each $m$ ($m=1,2$), from the joint probability distribution $\{p_{is}^{(m)}\}$, we can define two other joint probability
distributions $\{g_{is}^{(m)}\}$ and $\{q_{is}^{(m)}\}$, by defining the conditional probabilities as
$g_{s|i}^{(m)} = q_{s|i}^{(m)} = p_{s|i}^{(m)} = \bra{i_m} M_s \ket{i_m}$ and the prior probabilities as
$g_i^{(m)} = \frac{p_1^{(m)} - p_i^{(m)}}{dp_1^{(m)} -1}$ and $q_i^{(m)} = \frac{1}{d}$.
It is then obvious that
$q_i^{(m)} = \frac{1}{dp_1^{(m)}} p_i^{(m)} +(1-\frac{1}{dp_1^{(m)}}) g_i^{(m)} $ and
$q^{(m)}(s)= \sum_{i=1}^d q_i^{(m)} q_{s|i}^{(m)} = \frac{1}{dp_1^{(m)}} p^{(m)}(s) +(1-\frac{1}{dp_1^{(m)}}) g^{(m)}(s) $.
Therefore

\bgeqn
\frac{1}{dp_1^{(m)}} I_m &=& \frac{1}{dp_1^{(m)}} \left\{ H\{ p^{(m)}(s);s \} -\sum_{i}^d p_i^{(m)} H\{ p_{s|i}^{(m)};s \} \right\} \nonumber \\
&\leq& \frac{1}{dp_1^{(m)}} \left\{ H\{ p^{(m)}(s);s \} -\sum_{i}^d p_i^{(m)} H\{ p_{s|i}^{(m)};s \} \right\}  \nonumber \\
& & +(1-\frac{1}{dp_1^{(m)}} ) \left\{ H\{ g^{(m)}(s);s \} - \sum_{i=1}^d g_i^{(m)} H\{ g_{s|i}^{(m)} ;s\} \right\} \nonumber  \\
&=& \frac{1}{dp_1^{(m)}} H\{ p^{(m)}(s);s \} + (1-\frac{1}{dp_1^{(m)}} ) H\{ g^{(m)}(s);s \}   \nonumber \\
&& -
\sum_{i=1}^d \left(  \frac{1}{dp_1^{(m)}} p_i^{(m)} + (1-\frac{1}{dp_1^{(m)}} ) g_i^{(m)} \right) H\{ q_{s|i}^{(m)};s \} \nonumber \\
&\leq & H\{ q^{(m)}(s);s \} -  \sum_{i=1}^{d} q_i^{(m)} H\{ q_{s|i}^{(m)};s \}  \nonumber \\
&=& H\{ q^{(m)}_i;i \} -  \sum_{s} q^{(m)}(s) H\{ q_{i|s}^{(m)};i \}  \label{nonequal}
\edeqn
Here the inequalities follow from the concavity of Shannon entropy. As $q_i^{(m)} = \frac{1}{d}$, the right hand side of the above inequality is simply the mutual information via
the $m$th MUB when the prior probabilities are equal.  Using proposition 1, we immediately have (\ref{2mubgeneral1}).

{\bf Conjecture.} We have studied the problem numerically and found that when Alice uses different probabilities,
the inequality (\ref{2mubgeneral1}) does not provide a tight bound on the information on the two MUBs obtainable by Bob.
In fact, for all numerical examples, we find that unequal probabilities lead to lower information than that obtainable with
equal probabilities, which leads us to suggest as a conjecture, that proposition \ref{the11} holds also in the general case.

\bigskip

Suppose Alice has $M$ MUBs to choose from,
the sum of the information available to Bob via $M$ different MUBs is given by
\bgeq
I_{tot} = \sum_{m=1}^M I_m =\sum_m H\{ p^{(m)}_i;i \} - \sum_m \sum_s p^{(m)}(s) H\{ p^{(m)}_{i|s};i \} .
\label{Mmubsinfo}
\edeq

In analogy with our proposition 1, we have the following

\begin{theorem} \label{the21}
When Alice has $M$ MUBs to choose from and she transmits the basis states in each MUB with the same prior probability,
the information about the message sent via each MUB obtainable by Bob adds to $I_{tot}=\sum_{m=1}^{M} I_m$, and is restricted by the upper limits,
\bgeq
I_{tot}=\sum_{m=1}^{M} I_m \leq
M \log_2 \frac{d}{K}
- (K+1) \left( M- K \frac{d+M-1}{d} \right)
\log_2 \left( 1+ \frac{1}{K} \right) \leq
M \log_2 \frac{d+M-1}{M}  .
\label{dplus1mubs}
\edeq
with $K=\lfloor \frac{Md}{d+M-1} \rfloor$, and
\bgeq
I_{tot}=\sum_{m=1}^{M} I_m \leq
\frac{M}{2} \log_2 d
\label{dplus1mubs1003}
\edeq
which is stronger than (\ref{dplus1mubs}) when $M < \sqrt{d} +1$ and weaker than (\ref{dplus1mubs}) otherwise.
\end{theorem}

{\bf Proof.}
Since $p^{(m)}_i =1/d$, we have for the reverted protocol, in which Bob sends the state $\rho_s = M_s / \tr{M_s}$, that
$p^{(m)}(s)=\tr{M_s} /d \equiv p(s)$ does not depend on $m$, and therefore
we have a simpler form for $I_{tot}$,
\bgeq
I_{tot} = M \log_2 d - \sum_s p(s) \left( \sum_{m=1}^{M} H\{ p^{(m)}_{i|s};i \} \right) .  \label{equalmmub1}
\edeq
Since the prior probabilities are equal,
$p^{(m)}_{i|s}=\bra{i_m} \frac{M_s}{Tr M_s} \ket{i_m}$ can be viewed as the probability of
obtaining a result $i$ when the state $\frac{M_s}{Tr M_s}$ is measured by Alice with projective measurements in the
$m$th MUB.
For this dual situation (\ref{dplus1mubs}) follows directly from the entropic uncertainty inequality \cite{WYM08entunc}
\bgeq
\sum_{m=1}^{M} H\{ p^{(m)}_{i|s};i \} \geq
M \log_2 K
+ (K+1) \left( M- K  \frac{d+M-1}{d}  \right)
\log_2 \left( 1+ \frac{1}{K} \right) \geq  M \log_2   \frac{Md}{d+M-1} .
\label{the4100025}
\edeq
(\ref{dplus1mubs1003}) follows from the entropic uncertainty inequality \cite{BW07,WYM08entunc}
\bgeq
\sum_{m=1}^{M} H\{ p^{(m)}_{i|s};i \} \geq
\frac{M}{2} \log_2 d  .
\label{the4100778}
\edeq
This completes the proof of proposition \ref{the21}.

When $d$ is a power of a prime, it is known that $d+1$ MUBs exist. Suppose Alice uses
all the $d+1$ MUBs, i.e., $M=d+1$, by expansion of the logarithm,
we can show that the quantity on the right hand side of (\ref{dplus1mubs})
lies between $d-1$ and $d$, and it is strictly less than $d$ for any $d \geq 2$. Hence when $d$ is a power of a prime
and $M=d+1$, we have $I_{tot} < d$, or
equivalently,
\bgeq
\frac{1}{d+1} \sum_{m=1}^{d+1} I_m < \frac{d}{d+1} <1.
\edeq
The average mutual information per basis for the case of equal prior probabilities is less than $\frac{d}{d+1}$ bit when
$d$ is a power of a prime and $M=d+1$.

\medskip

When the prior probabilities are not equal, we can prove the following weaker results.

\begin{theorem}  \label{the22}
When Alice has $M$ MUBs to choose from and she transmits the basis
states in each MUB with different prior probabilities, the
information about each MUB obtainable by Bob add to
$I_{tot}=\sum_{m=1}^{M} I_m$, restricted by the upper limits, \bgeqn
\sum_{m=1}^{M} \frac{I_m}{d p^{(m)}_1} &\leq& M \log_2 \frac{d}{K} -
(K+1) \left( M- K \frac{d+M-1}{d} \right) \log_2 \left( 1+
\frac{1}{K} \right) \nonumber \\
&\leq& M \log_2 \frac{d+M-1}{M}, \label{eq2nonequal1}  \\
I_{tot} &\leq& d p_{max} \cdot \left\{
M \log_2 \frac{d}{K}
- (K+1) \left( M- K \frac{d+M-1}{d} \right)
\log_2 \left( 1+ \frac{1}{K} \right)
\right\} \nonumber \\
&\leq& d p_{max} \cdot \left\{ M \log_2 \frac{d+M-1}{M} \right\}
\label{eq2nonequal} \edeqn with $K=\lfloor \frac{Md}{d+M-1}
\rfloor$, and \bgeqn \sum_{m=1}^{M} \frac{I_m}{d p^{(m)}_1} \leq
\frac{M}{2} \log_2 d ,
\label{eq2nonequal10101}  \\
I_{tot} \leq d p_{max} \cdot \left\{
\frac{M}{2} \log_2 d
\right\} .
\label{eq2nonequal0202}
\edeqn
\end{theorem}
Here $p_{max}=max\{p^{(1)}_1, p^{(2)}_1, \cdots,p^{(d+1)}_1\}$, (\ref{eq2nonequal1}) is stronger than (\ref{eq2nonequal}).
(\ref{eq2nonequal10101}) and (\ref{eq2nonequal0202}) become stronger than (\ref{eq2nonequal1}) and (\ref{eq2nonequal})
when $M < \sqrt{d} +1$ and weaker otherwise.
The proof is similar to the proof of proposition \ref{the12}, and proposition \ref{the22} follows from
(\ref{nonequal}) and proposition \ref{the21}.

We have also studied the case of more than two MUBs numerically and as in the case of two MUBs, we find in all cases,
that the stronger inequality (\ref{dplus1mubs}), proven for equal probabilities, is observed also for different
probabilities, and we hence suggest the following conjecture.

\noindent
{\bf Conjecture.}
The sum over all $M$ MUBs of the information about the states sent in each MUB,
obtained by any general measurement procedure, has the upper limit (\ref{dplus1mubs}).

To our knowledge, the upper bounds, proven and conjectured above are  new results.
To arrive at our propositions and conjectures, we used results on the lower bound of
the entropic uncertainty in a ``reverted protocol".
There is a rich literature on the entropic uncertainty, and we imagine that our use of the reverted protocol, where Bob sends a measurement state to Alice, who performs projective measurements, may be
useful to obtain further connections between entropic uncertainty in one protocol and the available information
encoded via different bases in the other.

\section{Special results for qubits}

Let us now consider the special, but very important, case of a qubit ($d=2$), for which we have 3 mutually unbiased bases, corresponding, e.g., to the three pairs of eigenstates of the $X$, $Y$, and $Z$ spin components of a spin $1/2$ particle, or to polarization states of a photon, linearly polarized  along ($0^o,90^o$) or along ($45^o,135^o$) directions or circularly polarized, in the plane orthogonal to the direction of propagation.

The following corollary follows directly from propositions \ref{the21} and \ref{the22} as a special case.

\begin{corollary}. \label{coro1}
(a) When the prior probabilities are equal, the information available on the messages sent by a qubit in three MUBs obeys the inequality
\bgeq
I_1 + I_2 +I_3 \leq 1 . \label{3mubqubit3}
\edeq
(b) When the prior probabilities are not equal, the following inequalities apply (where we recall that the probabilities are ordered and $p^{(m)}_1$ is the largest probability within the  $m$th basis),
\bgeqn
\frac{I_1}{2 p^{(1)}_1} +\frac{I_2}{2 p^{(2)}_1} +\frac{I_3}{2 p^{(3)}_1}  &\leq& 1 , \\
I_1 + I_2 + I_3 &\leq& 2 p_{max} .
\edeqn
Here $p_{max}=max\{ p^{(1)}_1, p^{(2)}_1, p^{(3)}_1 \} $.
\end{corollary}

For the case of equal prior probabilities, (\ref{3mubqubit3}) is closely related to the result of
the dual problem in \cite{hall95} (eq. (9)) via a ``reverted protocol".

It is an interesting observation, that restrictions within the measurements available to Bob, affect the results, and,
e.g., if Bob is restricted to perform measurements with only two outcomes, we can prove that the tighter inequality
(\ref{3mubqubit3}) holds for the case of unequal probabilities,
$p^{(m)}_1=(1+\delta_m)/2,\ p^{(m)}_2=(1-\delta_m)/2$, chosen by Alice.
To prove this, we suppose without loss of generality that the three MUBs are given by the eigenstates of $\sigma_x$, $\sigma_y$ and $\sigma_z$.  Bob's two measurement elements are denoted by $M_{\pm} = \frac{1}{2} ( (1\pm R_0) I \pm \overrightarrow{R} \cdot \overrightarrow{\sigma} )$, where $|R_0|+R \leq 1$ as $M_{\pm} \geq 0$ ($R=|\overrightarrow{R}|$).
Using the expansion $\ln (1+ a) = \sum_{n=1}^{\infty} \frac{(-1)^{n-1}}{n} a^n$ for $-1< a \leq 1$, we get from a direct calculation
\bgeqn
I_x \ln 2&=&\sum_{n=1}^\infty \frac{1}{2n(2n-1)}
\left( \frac{1+\delta_x}2(R_0+R_x)^{2n}+\frac{1-\delta_x}2(R_0-R_x)^{2n}-(R_0+\delta_x
R_x)^{2n}\right)\cr
&=&\sum_{n=1}^\infty
\frac{1}{2n(2n-1)}\left(\sum_{k=0}^{n-1} (\begin{array}{c}
                                          2n \\
                                          2k
                                        \end{array})
 R_0^{2k}R_x^{2n-2k}(1-\delta_x^{2n-2k})+
\delta_x\sum_{k=1}^{n-1} (\begin{array}{c}
                                          2n \\
                                          2k-1
                                        \end{array})R_0^{2k-1}R_x^{2n-2k+1}(1-\delta_x^{2n-2k})\right)\cr
&\leq& \sum_{n=1}^\infty
\frac{1}{2n(2n-1)}\left(\frac{R_x^2}{R^2} \sum_{k=0}^{n-1}(\begin{array}{c}
                                          2n \\
                                          2k
                                        \end{array})R_0^{2k}R^{2n-2k}+
\frac{R_x^2}{R^2} \sum_{k=1}^{n-1}(\begin{array}{c}
                                          2n \\
                                          2k-1
                                        \end{array})|R_0|^{2k-1}R^{2n-2k+1}\right) \\ &=& \frac{R_x^2}{R^2} \sum_{n=1}^\infty \frac{1}{2n(2n-1)} \left( (|R_0|+R)^{2n} -R_0^{2n} -2n |R_0|^{2n-1}R \right) \cr &\le&  \frac{R_x^2}{R^2} \sum_{n=1}^\infty \frac{1}{2n(2n-1)} =  \frac{R_x^2}{R^2} \ln 2  \nonumber
\edeqn
Similar inequalities hold for $I_y$ and $I_z$, so we immediately get $I_x+I_y+I_z \leq 1$.

A numeric search has been performed for over 300 million cases. For
each case the probability distributions for the X, Y and Z basis
states are randomly generated, and Bob's measurement protocol is
also randomly generated with 2 to 16 rank-one POVM elements. Fig. 2
shows the values attained by the sum $I_1+I_2+I_3$. The prior
probabilities used by Alice are not equal, and the data are
organized horizontally according to the largest probability
$p_{max}$. This extensive numeric search supports the conjecture
that the tighter upper bound (\ref{3mubqubit3}) holds in the general
case. The figure also shows that when Bob uses only rank-one POVM
elements for the measurement, there is a non-trivial lower bound for
the information for each fixed $p_{max}$ \cite{remark1} (for equal
probabilities, in the left side of the plot, this minimum
information $I_{tot}= 0.768 $ is obtained by measuring the spin
along the direction $(1,1,1)$, when Alice uses the MUBs along the X,
Y and Z axes, in agreement with (26) in \cite{sanchez-ruiz95}) and (23) in \cite{sanchez93}.

\begin{figure}
\begin{center}
\includegraphics[angle=0,height=2in]{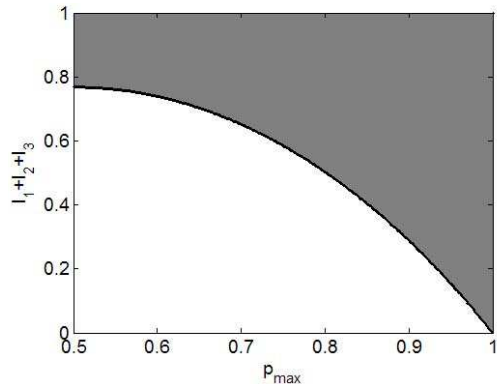}
\end{center}
\caption{The shaded region indicates the values attained by the sum $I_1+I_2+I_3$,
obtained by general POVM measurements with 2 to 16 rank-one members $M_s$. The region is identified by
randomly choosing the prior probabilities for the X, Y and Z basis states and randomly simulating
the measurements ($p_{max}$ is used as coordinate
axis in the plot).}
\end{figure}

\section{Complementarity of information available to different
observers via different MUBs: Security in QKD}
\nopagebreak

The information upper bounds proven above have applications to quantum key distribution, since they limit the information available to a potential eavesdropper about the communication shared between the legal users, Alice and Bob, of a transmission channel.

\subsection{BB84 protocol}

We consider first the BB84 protocol \cite{BB84} as an example.
Assume, that in her implementation of the BB84 protocol, Alice chooses either the Z or the X basis, and that she prepares her qubit in one of two basis states in either basis with equal
probability, before sending it to Bob.

An eavesdropper, Eve, may intercept the qubit before it reaches Bob. However,
the most general action of the receiver Bob and an eavesdropper, Eve, is accounted for by describing them as a single,
joint receiver, capable of carrying out general measurements on the states sent by Alice (Fig. 3). With this description,
viewing Bob and Eve
together as a single observer, BE, who performs measurements assuming no knowledge of which basis is used by Alice,
the inequalities discussed above can be used directly.

From proposition 1, we know \cite{remark2}
\bgeq
I_z( A: BE ) + I_x (A:BE) \leq 1  \label{sec4n1}
\edeq
where $I_{z(x)} (A:BE)$ is the information that Bob and Eve obtains about Alice's transmission in the Z (X) basis,
without knowing which basis is used by Alice.
Since the mutual information cannot increase if we discard a subsystem, we have,
\bgeqn
I_z(A:BE) &\geq & I_z (A:B) \\
I_x(A:BE) &\geq & I_x (A:E) .
\edeqn
So it follows that
\bgeq
I_z(A:B) + I_x(A:E) \leq 1 ,   \label{bb8411}
\edeq
and similarly we have
\bgeq
I_x(A:B) + I_z(A:E) \leq 1 . \label{bb8422}
\edeq

Even with a "helpful" Bob, who cooperates with Eve to let her extract the maximally allowed $Z$-information ($X$-information), while he reserves the  $X$-information ($Z$-information) for himself, the sum of these amounts of mutual information obeys our simple upper bound. Therefore by checking the amount of information sent
through one axis Alice and Bob can determine the maximum information available to a
hypothetical Eve, intercepting the
information via the other axis. By checking via both axes for a given amount of test qubits, they can thus ascertain a maximum level of eavesdropping on the channel and apply privacy amplification to enable fully confident transmission.
It is often explained how the security of QKD is due to the complementarity of information sent through different MUBs, and the present analysis shows this quantitatively.

\begin{figure}
\begin{center}
\includegraphics[angle=0,height=0.8in]{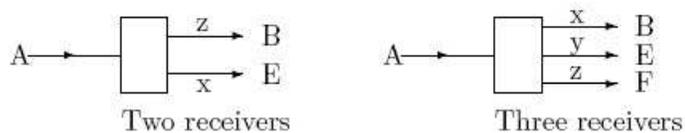}
\end{center}
\caption{\label{fig3} Illustration of complementarity of information available to different observers via
different MUBs. In the two-receiver scheme (BB84 protocol), the X information obtained
by Eve decreases the available Z information for Bob. In the three-receiver scheme,
Eve's information obtained about one basis and Fred's information obtained about another MUB both
decrease the available information for Bob about the third MUB.}
\end{figure}

\subsection{Six state protocol}

We extend the BB84 protocol to the case where Alice chooses between three qubit bases, $Z$, $X$ or $Y$,
and prepares the qubit in one of two basis states of her chosen basis with equal prior probability.
To investigate the possibilities of eavesdropping on this six-state protocol \cite{Bruss98,BG98}, we consider now three receivers: Bob, Eve
and Fred, where the eavesdroppers Eve and Fred could be one and the same person (Fig. 3).
Treating first Bob, Eve and Fred as a single receiver, capable of performing the most general measurements
on the quantum system, and noting again that information does not increase upon discarding subsystems,
from corollary \ref{coro1}, we get
\bgeq I_x (A:B) + I_y(A:E) + I_z (A:F) \leq 1 . \edeq
Eve's attempt to acquire $Y$-information and Fred's attempt to acquire
$Z$-information will add up to the decrease in the available
$X$-information for Bob.

\subsection{Larger dimensional protocols}

Similar inequalities can be obtained when Alice sends qudits to Bob instead of qubits. Apart from being interesting in its own right, we note that this may constitute a useful starting point for the analysis of upper bounds of the information available by collective attacks (on several subsequently emitted qubits) on the BB84 and the 6-state protocols.

\section{Conclusion}
\nopagebreak
\bigskip

In conclusion, we have quantitatively investigated the
complementarity of information obtainable via different mutually
unbiased bases (MUBs), when the sender, Alice, has more than one MUB
at her disposal to encode her messages and the receiver, Bob, has no
prior knowledge of her choice of basis. We obtain various upper
bounds on the sum of information available to Bob encoded by Alice
via $M$ different MUBs.

We prove that the sum of Bob's available information encoded via two different
MUBs has an upper bound of $ \log_2 d$ when the prior probabilities of
Alice's messages are all equal to $1/d$ for each MUB.
We also prove weaker upper bounds for the case of non-equal prior probabilities.
We give and prove upper bounds of
the sum of Bob's available information encoded via
$M$ MUBs when the prior probabilities of Alice's messages are all equal to $1/d$.
Weaker upper bounds for the general case
of non-equal prior probabilities are also obtained and proved.
Numerical searches indicate that the tighter upper bounds proven for equal probabilities also apply in the general case of unequal probabilities used by Alice, but only for qubits and under a restriction on the measurements available to Bob, we have found rigid proof of such a conjecture.

Using the complementarity of information sent via different MUBs, we
derive quantitative inequalities to show that there is also a
complementarity of the information obtainable by different
observers, when each of them tries to gain information encoded via a
different MUB. This latter complementarity is the essence of the
security of QKD protocols: the eavesdropper's knowledge of the
information sent via one MUB inevitably decreases the information
sent to another observer via another MUB.

\section*{Acknowledgments}
The authors wish to thank Uffe V. Poulsen and the MOBISEQ network under the Danish Natural Science Research Council
for helpful discussions,
and Li Yu for drawing our attention to an unpublished alternative proof of (\ref{bb8411}) and (\ref{bb8422})
\cite{griffithsunp}.
S. W. and S. Y. also wish to acknowledge support from the NNSF of China (Grants No. 10604051 and No. 10675107),
the CAS, and the National Fundamental Research Program.


\begin{thebibliography}{99}
\bibitem{remark1}
In terms of upper bound, it is sufficient for Bob to use only rank-one POVM elements for the measurement,
since the mutual information will not decrease when Bob's POVM elements are refined into rank-one POVM elements.
However, the non-trivial lower bound in the figure exists only when Bob's POVM elements are rank-one.
\bibitem{remark2}
The inequality (\ref{sec4n1}) is closely related to (78) in
\cite{hall97} with a specified choice of the signal states, and we
suspect that our analysis may also be generalized to derive results
similar to (77) in \cite{hall97} for measurements on correlated
pairs of particles.
\end{thebibliography}
\end{document}